\documentclass[lettersize,journal]{IEEEtran}
\usepackage{amsmath,amsfonts}
\usepackage{algorithmic}
\usepackage{algorithm}
\usepackage{array}
\usepackage[caption=false,font=normalsize,labelfont=sf,textfont=sf]{subfig}
\usepackage{textcomp}
\usepackage{stfloats}
\usepackage{url}
\usepackage{verbatim}
\usepackage{graphicx}
\usepackage{cite}
\usepackage{xcolor}
\usepackage{siunitx}
\usepackage{booktabs}
\usepackage[para]{threeparttable}
\usepackage{multirow}
\usepackage[normalem]{ulem}
\usepackage{siunitx}
\usepackage{tabularx}

\usepackage{url}

\usepackage{breakurl}
\usepackage[breaklinks,hidelinks]{hyperref}

\makeatletter
\def\ps@IEEEtitlepagestyle{%
  \def\@oddfoot{\mycopyrightnotice}%
  \def\@oddhead{\hbox{}\@IEEEheaderstyle\leftmark\hfil\thepage}\relax
  \def\@evenhead{\@IEEEheaderstyle\thepage\hfil\leftmark\hbox{}}\relax
  \def\@evenfoot{}%
}
\def\mycopyrightnotice{%
  \begin{minipage}{\textwidth}
  \centering \scriptsize
  This work has been submitted to the IEEE for possible publication. Copyright may be transferred without notice, after which this version may no longer be accessible.
  \end{minipage}
}
\makeatother

\begin{document}

\title{A Spiking Neural Network Decoder for Implantable Brain Machine Interfaces  and its Sparsity-aware Deployment on RISC-V Microcontrollers }


\author{Jiawei Liao,~\IEEEmembership{Graduate Student~Member,~IEEE,}
Oscar Toomey,~\IEEEmembership{Member,~IEEE,} \\
Xiaying Wang,~\IEEEmembership{Member,~IEEE,}
Lars Widmer, 
Cynthia A. Chestek,~\IEEEmembership{Senior Member,~IEEE,} \\
Luca Benini,~\IEEEmembership{Fellow,~IEEE,}
Taekwang Jang,~\IEEEmembership{Senior Member,~IEEE}


\thanks{Jiawei Liao, Oscar Toomey, Xiaying Wang, Lars Widmer, Luca Benini and Taekwang Jang are with the Department of Information Technology and Electrical Engineering, ETH Zürich, 8092 Zürich,
Switzerland (Email: liaoj@iis.ee.ethz.ch). Xiaying Wang is also with the Swiss University of Traditional Chinese Medicine, 5330 Bad Zurzach, Switzerland.

Cynthia A. Chestek is with the Department of Biomedical Engineering
and Electrical Engineering and Computer Science, University of Michigan, Ann Arbor, MI 48109 USA.}}



\maketitle

\begin{abstract}
Implantable Brain-machine interfaces (BMIs) are promising for motor rehabilitation and mobility augmentation, and they demand accurate and energy-efficient algorithms. In this paper, we propose a novel spiking neural network (SNN) decoder for regression tasks for implantable BMIs. The SNN is trained with enhanced spatio-temporal backpropagation to fully leverage its capability to handle temporal problems. The proposed SNN decoder outperforms the state-of-the-art Kalman filter and artificial neural network (ANN) decoders in offline finger velocity decoding tasks.

The decoder is deployed on a RISC-V-based hardware platform and optimized to exploit sparsity. The proposed implementation has an average power consumption of 0.50mW in a duty-cycled mode. When conducting continuous inference without duty-cycling, it achieves an energy efficiency of 1.88$\mathbf{\mu}$J per inference, which is 5.5X less than the baseline ANN. Additionally, the average decoding latency is 0.12ms for each inference, which is 5.7X faster than the ANN implementation. \end{abstract}


\section{Introduction}
\IEEEPARstart{B}{rain-machine} interfaces (BMIs) bridge neurons in the brain with external electronic devices. They have become increasingly popular in both academia and industry as a promising approach to assist individuals with paralysis and amputations in regaining or enhancing their motor functions. An early study in 2004 conducted a human clinical trial to showcase prosthetic devices with an implanted BMI~\cite{hochberg_neuronal_2006}. A BMI proposed in~\cite{willett_high-performance_2021} was able to decode handwriting movement intentions at a speed close to typical smartphone texting. In a more recent study, researchers demonstrated a BMI system that bridges the brain and spinal cord to treat a patient with spinal cord injury. By recording and processing the brain's neural signals to stimulate neurons in the spine, this new BMI system has restored the person's ability to walk~\cite{lorach_walking_2023}.

A BMI system typically consists of three main components: data acquisition, data processing, and stimulation or actuation. Both non-invasive and invasive methods are viable choices for acquiring data from the brain. Although non-invasive methods, such as EEG, reduce risks for patients, invasive methods such as iEEG or ECoG usually outperform the non-invasive methods in terms of precision thanks to their higher signal-to-noise ratio and finer temporal-spatial resolution~\cite{rapeaux_implantable_2021}. Despite the advantages, implanted BMIs are often tethered using wires for the transfer of power and the large amount of data necessary for accurate decoding~\cite{hochberg_neuronal_2006,willett_high-performance_2021}, resulting in a higher risk of scar tissue formation and infection. One promising approach to reduce the damage caused by the wires is neural dust~\cite{seo_wireless_2016}, featuring wireless power and data transfer to and from the miniaturized and distributed devices~\cite{lim_269_2020}. However, this wireless approach faces stringent requirement, arising from two factors: 1) limited battery life and transferable power, and 2) the risk of tissue damage caused by the heat generated by the implant. Among the components of a BMI system, the data transceiver is usually the most power-hungry element, given its role in handling vast amount of data acquired from high-channel recording circuits. To mitigate these challenges, edge computing emerges as a viable solution for BMI systems. By processing data directly within the sensor system, it reduces data transfer power. This indicates the growing importance of developing energy-efficient neuronal decoding algorithms within the BMI field.

\IEEEpubidadjcol

Spiking neural networks (SNNs), inspired by biological neurons, are promising for energy-efficient inference. Neurons in SNNs employ single-bit spikes to transfer information between layers, unlike neurons in conventional artificial neural networks (ANNs) that generally use multiple bits. This property allows for efficient hardware implementation. Moreover, SNNs possess inherent temporal memory, allowing them to extract temporal features and making them well-suited for time-series regression tasks such as motor function decoding. Another beneficial property of SNNs lies in the firing behavior of neurons, which occurs only occasionally and asynchronously. This leads to data sparsity that can be harnessed to achieve more energy-efficient computation. Thanks to their energy-efficient properties, SNNs have been extensively explored for classification tasks including image classification~\cite{wu_spatio-temporal_2018,zheng_going_2020} and gesture recognition~\cite{donati_discrimination_2019,baracat_neuromorphic_2024}, yet more studies are necessary to assess their usage in time-series regression tasks.

Although SNNs show promise for energy-efficient implementation, several challenges remain before this potential can be fully realized. Firstly, SNNs have a more complex data flow than conventional ANNs. In contrast to typical fully connected or convolution networks, SNNs require additional handling of temporal memory. Secondly, despite the natural temporal data sparsity of SNNs, effectively exploiting the sparsity on resource-constrained hardware remains a significant challenge. The sparsity lies in feature maps varying from one inference to another, leading to irregular memory access patterns. Thirdly, many SNN implementations require multiple inference cycles to form the temporal dimension for a single input data due to the rate encoding, increasing energy consumption, and processing time~\cite{narayanan_spinalflow_2020}. Therefore, despite the benefits of single-bit operation and sparsity of SNNs, their practical application has been limited to neuromorphic hardware specifically designed for this purpose with inherent limitations in flexibility and adaptability~\cite{malcolm_comprehensive_2023}. On the other hand, recent general-purpose edge computing platforms have proven to be energy efficient and suitable for BMI applications~\cite{wang_mi-bminet_2024,wang_sub-100_2021}. Exploring SNN deployment on these platforms is beneficial since they are more versatile, affordable, and widely accessible. They are optimal for algorithmic and implementation explorations before the final deployment on neural implants, as suggested by~\cite{liu_edge_2021}.

\begin{figure*}[!t]
 \centering\includegraphics[width=\linewidth]{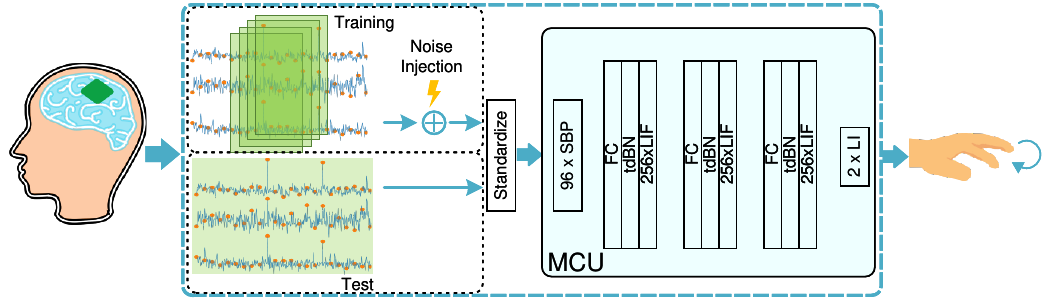}
 \caption{Application scenario and data flow of the proposed design. The neural signal is obtained from the motor cortex region of a brain before the SBP feature is extracted. During training, the input data is prepared with a sliding window containing 10 time steps with 9 steps overlapping. Then, noise is injected before the data is normalized. In test mode, all the data is streamed into an SNN neural decoder deployed on an MCU that predicts the finger velocity. }
 \label{fig:dataflow}
\end{figure*}

In this work, we propose a neural decoding algorithm and deployment methodology for a low-complexity SNN, trained with an enhanced backpropagation method, aimed at the offline open-loop finger velocity prediction for implantable BMIs, as depicted in Fig.~\ref{fig:dataflow}.
Previous research in ~\cite{liao_energy-efficient_2022} presents an SNN-based finger velocity decoder, reporting decoding accuracy and computational complexity. The achieved decoding coefficients were 0.745 and 0.582 for the two given datasets. Building on~\cite{liao_energy-efficient_2022}, this work introduces new regularization techniques, yielding improved accuracies of 0.782 and 0.624. Furthermore, this paper delves into the implementation details, such as the model, training method, and the entire hardware deployment flow for inference, including the quantization and optimization for the deployment on the energy-efficient hardware platform.

The main contributions of this paper are summarised below:
\begin{itemize}
    \item We propose an open source\footnote{https://github.com/liaoRichard/SNN-for-Finger-Velocity-iBMI (To be updated on acceptance)} and low-complexity SNN for continuous-time finger velocity decoding for low-power implantable BMIs. 
    \item We demonstrate an SNN backpropagation training strategy enhanced with neuron reset-by-subtraction, trainable decay factors, and noise injection to achieve a higher accuracy than the state-of-the-art KF~\cite{nason_real-time_2020} and ANN decoders~\cite{willsey_real-time_2022}.
    \item The proposed design is quantized with quantization-aware training and deployed to a RISC-V-based GAP9 microcontroller (MCU) for inference, resulting in a memory footprint saving of around 4X. The deployment is optimized to leverage SNN sparsity to enhance performance and efficiency. 
    \item The implementation achieves an energy per inference of \SI{1.88}{\micro\joule} for continuous inference, 5.7X less than the baseline ANN and 2.1X less than the baseline SNN implementation that does not exploit sparsity. The average inference latency is only \SI{0.12}{\milli\second}, 5.5X and 1.8X faster than the conventional ANN and SNN baseline, respectively. 
\end{itemize}

To the best of our knowledge, our work is the first sparsity-aware implementation of an SNN on a general-purpose edge computing platform for neural implants.


\section{Related Works}

\begin{table*}
    \centering
    \caption{Comparison table}
    \label{tab:comp_table}
    \begin{threeparttable}
    \setlength{\tabcolsep}{3pt}
    \begin{tabular}{@{}lrrrrrrrr@{}l} 
    \hline
         &  This Work & BioCAS24~\cite{leone_spiking_2023}  & TBioCAS22~\cite{an_power-efficient_2022} & TBioCAS16~\cite{chen_128-channel_2016} & ISSCC24~\cite{shaeri_333_2024} & Frontiers16~\cite{boi_bidirectional_2016}   &ISCAS23~\cite{boretti_event-based_2023} & IEEE Sens.J24~\cite{wang_mi-bminet_2024} \\ 
         \hline
         Type&  MCU&  FPGA&  ASIC &ASIC &  ASIC&ASIC  &MCU &MCU\\ 
         
 Task
 & \begin{tabular}{@{}r@{}}BMI Finger \\ Movement\end{tabular} 

 & \begin{tabular}{@{}r@{}}BMI Finger \\ Movement\end{tabular} 
 & \begin{tabular}{@{}r@{}}BMI Finger \\ Movement\end{tabular}
  &\begin{tabular}{@{}r@{}}BMI Finger \\ Movement\end{tabular} 
  & \begin{tabular}{@{}r@{}}BMI Hand \\ Writing\end{tabular} 
 &\begin{tabular}{@{}r@{}}BMI Object \\ Control\end{tabular}
 &\begin{tabular}{@{}r@{}}Digit \\ Recognition\end{tabular} 
 & \begin{tabular}{@{}r@{}}BMI \\ Motor Imagery\end{tabular}
\\ 
         Feature&  SBP&  MUA &  SBP &Spike rate &  Spike rate &Spike rate  &Spike rate & EEG\\ 
         Algorithm&  SNN&  SNN &  SSKF &ELM &  DNC + LDA &SNN  &SNN & CNN\\
 Output Type& Regression& Regression & Regression &12 classes & 31 classes&4 clases  &10 clases & 2 classes\\ 
 Accuracy 
 &  \begin{tabular}{@{}r@{}} 0.782 corr \\ 0.627 corr\end{tabular} 
 &  0.84 corr \tnote{c} 
 &  \begin{tabular}{@{}r@{}} 0.601 corr\tnote{d} \\ 0.459 corr\tnote{d}\end{tabular} 
 &  \begin{tabular}{@{}r@{}} 99.3\%\tnote{c} \end{tabular} 
 &  \begin{tabular}{@{}r@{}} 90.8\%\tnote{c} \end{tabular} 
 &  \begin{tabular}{@{}r@{}} 50\% - 70\%\tnote{c} \end{tabular} 
 &  \begin{tabular}{@{}r@{}} 97.2\% \tnote{c}\end{tabular} 
 &  \begin{tabular}{@{}r@{}} 82.51\% \tnote{c}\end{tabular} \\
 Power&  0.50 mW&  56.4 mW &  0.588 mW \tnote{a} &0.4 uW\tnote{b}  &  0.22 mW&4 mW &99.5 mW & 10.17 mW\\ 
 Latency& 0.12 ms& 0.3 ms & $<$ 1 ms &-& -& $<$ 100 ms  &0.04 ms & 2.95 ms\\ 
 Energy/infer. & 1.88 uJ& 56.4 uJ & -&- & -&-  &4.9 uJ & 30uJ\\ 
 \hline
    \end{tabular}

    \begin{tablenotes}
\footnotesize
\item[a] Feature extraction power included
\item[b] MCU power not included
\item[c] Not same tasks or datasets, cannot be used for direct comparison
\item[d] Reproduced results, assuming SSKF has same accuracy as KF

\end{tablenotes}
\end{threeparttable}
\end{table*}

\subsection{Neural decoding algorithms}
Neural decoding algorithms translate brain activities into signals that can be used to operate actuators based on the user's intention. Linear decoders, such as linear regression, linear discriminant analysis, and variants of Kalman filters have been developed to perform arm and hand control~\cite{collinger_7_2013,hotson_individual_2016,ajiboye_restoration_2017,nason_real-time_2020,hochberg_reach_2012}. Linear decoders require little computation and can be implemented in highly energy-efficient hardware. However, linear decoders can only achieve moderate accuracy. Nonlinear decoders, including recurrent neural networks~\cite{sussillo_recurrent_2012,hosman_bci_2019} and feed-forward neural networks~\cite{glaser_machine_2020,willsey_real-time_2021} have received recent attention. While neural networks are powerful, they come at the cost of high computational complexity, implying high energy consumption in hardware implementations~\cite{sze_efficient_2017}. Therefore, it is essential to develop decoding algorithms that can achieve high accuracy while consuming low energy. To this end, SNNs emerge as promising candidates. Our 
proposed SNN achieves better accuracy compared with the state-of-the-art KF and ANN neural decoders in offline neural decoding tasks reported in table~\ref{tab:acc_comp}, which will be discussed in detail in section~\ref{sec: result}.

\subsection{SNN training and deployment}
There are three main methods for SNN training:

1) ANN-to-SNN conversion requires an ANN trained before converting it to an SNN. Typically, it relies on rate coding and tries to mimic the computation of the ANN by setting the parameters of the SNN to correlate the spiking neurons' firing rates with the neurons' activation of the original ANN~\cite{rueckauer_conversion_2017}.  In the same vein, \cite{dethier_spiking_2011} introduced an SNN decoder converted from a Kalman filter decoder. This approach has shown accuracy comparable to its ANN or KF counterparts but typically requires high spike counts due to the conversion process and rate approximation, resulting in high energy consumption and latency.

2) Unsupervised learning, which relies on local learning rules~\cite{masquelier_unsupervised_2007} such as spike-timing-dependent-plasticity (STDP). This is a biologically plausible approach, but without global supervision, it has not yet achieved state-of-the-art performance in terms of accuracy and energy efficiency.

3) SNN backpropagation such as spatio-temporal-back-propogation (STBP)~\cite{wu_spatio-temporal_2018,zheng_going_2020}. This approach establishes an error backpropagation path for gradient descent training by applying a surrogate function in the backward flow to approximate the derivative of the spike activity. It utilizes the temporal feature of the input and has demonstrated good performance in classification tasks. In this work, we construct the SNN using the SNN backpropagation method.

After training, SNNs are deployed. The key performance indicator for deployment is energy efficiency. Due to the irregular memory access pattern and more complex data flow, deploying SNNs efficiently can be challenging. FPGA accelerators~\cite{neil_minitaur_2014}, and analog~\cite{moradi_scalable_2018} or digital~\cite{frenkel_morphic_2019,akopyan_truenorth_2015,davies_loihi_2018} ASIC accelerators have been designed targeting efficient deployment of SNNs. There are only a few studies discussing SNN deployment on MCUs. An SNN is deployed on a low-cost MCU for digit classification task~\cite{boretti_event-based_2023}, which, while demonstrating feasibility, does so without detailed optimization or deployment strategies. This system reports a relatively high power consumption of \SI{99.5}{\milli\watt} and an energy efficiency of \SI{4.9}{\micro\joule} per inference. In contrast, our paper offers an in-depth optimization of SNN deployment on MCU, with special emphasis on exploiting sparsity.

\subsection{BMI hardware systems}
Table~\ref{tab:comp_table} provides a comparative analysis of our approach with previous works in the field of BMI hardware deployment and implementation. 

The works in~\cite{wang_mi-bminet_2024,chen_128-channel_2016,shaeri_333_2024,boi_bidirectional_2016} focus on implementing BMIs for classification tasks.
Specifically, \cite{wang_mi-bminet_2024} introduces a CNN deployed on a RISC-V-based MCU for motor imagery tasks, consuming tens of uJ per inference and showing the feasibility of applying modern MCUs to BMI applications. 
\cite{chen_128-channel_2016} discusses a mixed-signal ASIC for classifying finger movements using an extreme learning machine (ELM) algorithm, which consumes only \SI{0.4}{\micro\watt}. However, it should be noted that this measurement excludes the power of the MCU and peripherals. Moreover, the adaptability of ELM for BMI regression tasks requires further exploration.

\cite{leone_spiking_2023,an_power-efficient_2022} introduce BMI hardware for regression tasks. \cite{an_power-efficient_2022} describes an ASIC that supports feature extraction and decoding, maintaining a low power consumption of \SI{588}{\micro\watt} and a processing latency under \SI{1}{\milli\second} for closed-loop BMI tasks. However, the KF-based decoder in this system yields relatively lower accuracy than neural network-based decoders.

In a recent work by Leone, et al.~\cite{leone_spiking_2023}, an FPGA-based SNN decoder was introduced for finger movement regression tasks, with a latency of only \SI{0.3}{\milli\second}. 
However, this design has a power consumption of \SI{56.4}{\milli\watt} and energy per inference of \SI{56.4}{\micro\joule} which are significantly higher when compared to other neural decoders.

Compared to the prior works, our design demonstrates an SNN-based decoder and its sparsity-aware deployment on a flexible MCU with competitive decoding accuracy, latency of \SI{0.12}{\micro\second}, and energy efficiency of \SI{1.88}{\micro\joule\per inference} in BMI regression tasks.

\section{Algorithm}
\subsection{Input feature}

The proposed network adopts the spiking band power (SBP) introduced in ~\cite{nason_low-power_2020} as the input feature. SBP is a neural feature defined as an absolute average of a $\SI{300}-\SI{1000}{\hertz}$ band-pass-filtered signal. The advantage of the spiking band power feature lies in its lower data rate than the direct action potential acquisition, implying lower power for communication, while its value is still determined mainly by single neuron spikes. SBP has shown good performance for motor prediction tasks when embedded in various algorithms~\cite{nason_real-time_2020,willsey_real-time_2022}. SBP feature extractors have been implemented in both digital~\cite{an_power-efficient_2022} and analog~\cite{lim_269_2020} circuits consuming very low power. In this work, we propose a network receiving SBP features from 96 channels. 

\subsection{Neuron Model}
\label{sec: neuron model}
The proposed network uses a leaky-integrate-and-fire (LIF) neuron as its neuron model, which is one of the most commonly applied models in computational neuroscience. It is inspired by the biological neuron but implemented in a simplified form. As demonstrated in~\cite{izhikevich_which_2004}, the LIF neuron model shows a good compromise between cognitive capabilities and computational complexity, thus making it a suitable candidate for embedded platforms with limited resources. The action potentials are simplified to events and neural dynamics, which are governed by two equations describing the membrane potential evolution and the spike generation mechanism~\cite{gerstner_neuronal_2014}. The membrane potential is defined by the differential equation 
\begin{equation}
\label{eq: lif_continous_time}
\tau \frac{du}{dt} = -u + i,
\end{equation}
where $u$ is the membrane potential, $i$ is the input current, and $\tau$ is the time constant for the membrane potential decay. 

To make the model more suitable for efficient computation, we can solve the equation \eqref{eq: lif_continous_time} with the forward Euler method, which gives us the discrete-time equation 
\begin{equation}
\label{eq:lif_iterative}
u(t+dt) = (1 - \frac{dt}{\tau})u(t) + \frac{dt}{\tau}i.
\end{equation}

 In the scenario of neural networks, the input $i$ is the weighted sum of the output from a previous layer and can be represented by 
 \begin{equation}
\label{eq:inp_current}
i^{l}(t) = W^{l}s^{l-1}(t) + B^{l},
\end{equation}
 where $W$ is the weight matrix, $B$ is the bias vector, and $l$ represents the layer number.

The neuron fires its output in accordance with its current membrane potential. This behavior is governed by the second half of the neural dynamics model, the spike generation mechanism,  represented by a simple Heaviside step function 
\begin{equation}
\label{eq:spike_fun}
s^{l}(t) = \begin{cases}
1   & u^{l}(t) - V_{th} \ge 0  \\
0   & u^{l}(t) - V_{th} < 0,
\end{cases}
\end{equation}
where s represents the single-bit output of a neuron, and Vth is the threshold voltage.

The membrane potential will reset to a new value after the neuron fires. Typically, when the membrane potential exceeds a threshold, the neuron fires and resets its membrane potential to zero, which means all information regarding recent input activity is lost after firing. However, we adjusted the reset scheme to subtract the threshold voltage ($V_{th}$) from the membrane potential when firing, as inspired by \cite{tan_improved_2021}. This means that the information from very high activities is not completely lost. The data above the threshold can influence the neuron’s decision at the next firing. This adaptation poses only a little hardware overhead, requiring only one subtraction operation per neuron per inference.

The equation \eqref{eq:lif_iterative} can be further simplified by using decay factors $\lambda = 1 - \frac{dt}{\tau}$ and merging the factor $\frac{dt}{\tau}$ into the weights of the input. After incorporating the spikes and resetting, the membrane potential can be modeled by 

\begin{equation}
\label{eq:mem_upd}
u^{l}(t) = \lambda (u^{l}(t-1) - s^{l-1}(t-1)V_{th}) + i^{l}(t).
\end{equation}

\begin{figure}[t]
 \centering\includegraphics[width=\linewidth]{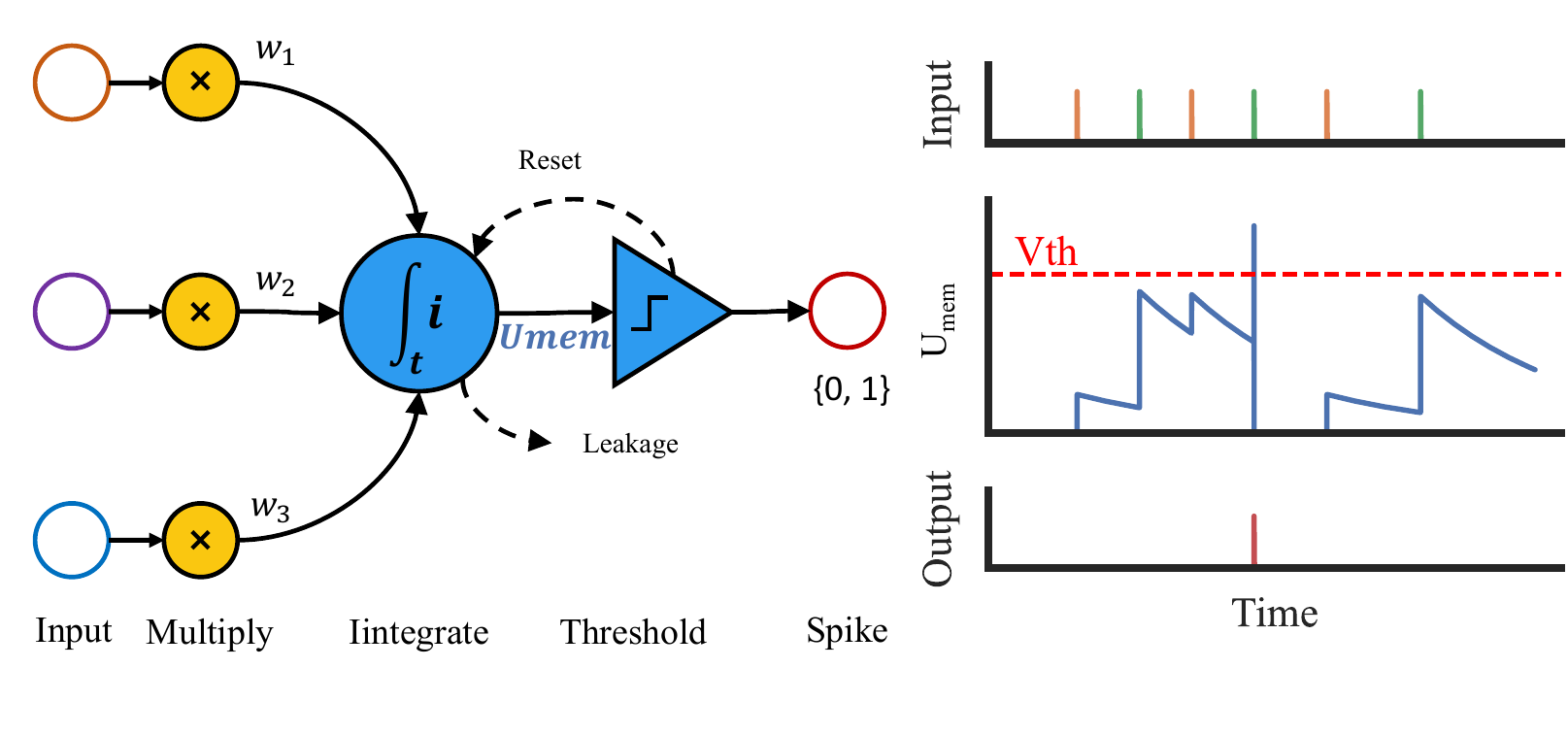}
 \caption{LIF neuron dynamics. Membrane potential $u$ keeps the internal state of a neuron. If the membrane potential is higher than a threshold voltage, then the neuron will fire and reset the membrane potential. Otherwise, the neuron keeps its membrane potential for the next step and does not fire.}
 \label{fig:neuron model}
\end{figure}

The behavior of our neural model follows \eqref{eq:inp_current}, \eqref{eq:spike_fun}, and \eqref{eq:mem_upd}, and it is illustrated in Fig.~\ref{fig:neuron model}.
The membrane potential $u$ keeps the internal state of a neuron. The membrane potential and the decaying factor $\lambda$ allow the neuron to partly retain its former information and inherently capture temporal dynamics.

\begin{figure}[t]
 \centering\includegraphics{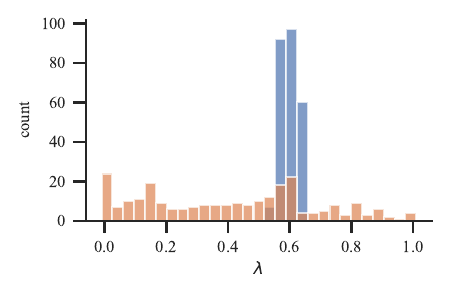}
 \caption{Decay factor $\lambda$ distribution before and after training. Tau is randomly initialized around 0.6. After training, $\lambda$ spread in the range between 1 and 0}
 \label{fig:tau_dist}
 
\end{figure}

The decay factor $\lambda$ determines the length of history that a neuron can remember. Larger $\lambda$ leads to slower membrane potential decay, while small values lead to fast decay. Different $\lambda$ values indicate the different lengths of history that a neuron can remember. Typically, a single $\lambda$ is employed across the whole network. However, in our proposed network, each neuron has its own individually trainable decay factor $\lambda$, allowing for varying sensitivities to historical data among neurons. In this way, we allow the network to capture more complex temporal dynamics. Fig.~\ref{fig:tau_dist} shows the distribution of the decay factor that spans the entire range between 0 and 1 after training. Introducing trainable decay factors increases the complexity of training. However, in the inference process, its overhead of storing the decay factor is negligible since the number of neurons is far fewer than the number of weights.

In contrast to the normal ANN implementation where the information between layers is carried by multi-bit real values, the information between spiking layers is encoded in 1-bit spikes. Therefore, one of the advantages of using an SNN instead of a conventional ANN is that the input current $i$ can be calculated as cheaper additions of weights instead of Multiply-Accumulate (MAC) operations due to the 1-bit input spikes. All hidden neurons fire only occasionally when the membrane potentials exceed the threshold, introducing a high degree of sparsity in the intermediate features. The reduced computational complexity and the high degree of sparsity are the two main sources of energy efficiency of the proposed SNN compared to the conventional ANNs. 

\subsection{Proposed Network Architecture}

\begin{figure}[t]
  \centering\includegraphics{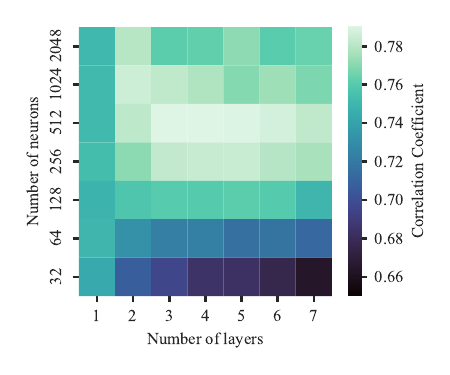}
 \caption{Topology sweep. The scope of this exploration includes the number of spiking layers and the number of neurons in each layer except for the output layer.}
 \label{fig:topo_sweep}
\end{figure}

In this work, we propose a fully-connected SNN for velocity regression. A conventional ANN is proposed in~\cite{willsey_real-time_2022} for the same task, where the input layer is a temporal convolution layer for feature extraction from multiple time steps, followed by three fully connected layers. However, the large number of neurons in the flattened output of the temporal convolutional layers leads to a high number of parameters and computations in their second layer.   

Instead of processing multiple time steps of inputs using a temporal convolution layer, we only pass inputs from a single time step in each inference. Then, we rely on the implicit recurrent nature of the neurons for temporal processing. As a result our model only uses 4 fully connected spiking layers. 
This choice was made to exploit an intrinsic property of SNNs, i.e., the membrane potential of each neuron acts as a memory element and stores temporal information from previous inputs. 

The input to the first layer of our proposed network are multi-bit SBP features from 96 channels. Hence, the input current is calculated as $i(t) = \sum_{n=0}^{M-1} w  SBP_n(t)$, rather than the  $i(t) = \sum_{i=0}^{M-1} w^{l} s^{l-1}(t)$ used in the other layers. Hence, for the first layer, it is not possible to replace MAC operations with additions.
The second and third layers are spiking layers formed by the LIF neurons described in section~\ref{sec: neuron model}. The output layer is composed of two leaky-integrate neurons predicting the velocity of two fingers. In contrast to some works converting the spikes back to the real value, we directly use the membrane potential as the output. Therefore, the neurons in the last layer are modeled by $u(t) = \lambda u(t-1) + i(t) $ without spike and reset mechanisms.

\begin{figure}[t]
 \centering\includegraphics[width=\linewidth]{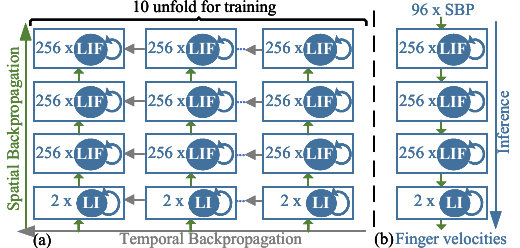}
 \caption{Proposed SNN unfolded for training (a) and in inference (b)}
 \label{fig:network_arch}

\end{figure}

The number of hidden layers and neurons in each layer is determined based on the parameter sweep shown in Fig.~\ref{fig:topo_sweep}. We choose a configuration consisting of 4 layers, each comprising 256 neurons except for the output layer, after making a trade off between the configuration size and the performance improvement. 

Between successive layers, batch normalization is implemented to improve convergence.  Training an SNN with a backpropagation approach requires unfolding the network, equivalent to expanding the network to more layers. This severely exacerbates the gradient vanishing or explosion problems. Unlike the conventional ANN, the SNN has an extra temporal dimension, and neurons firing too frequently or scarcely can degrade performance~\cite{zheng_going_2020}. Therefore, the scale of the threshold voltage and the input needs to be balanced. We employ the threshold-based batch normalization introduced in~\cite{zheng_going_2020} to address these issues. The threshold-dependent batch normalization can be described by 

\begin{equation}
\label{eq:tdBN}
y = \frac{V_{th}(x-E[x])}{\sqrt{Var[x]+\epsilon}}*\gamma + \beta.
\end{equation} 

In training, $E[x]$ and $Var[x]$ are calculated over the mini batches. In inference, we use the mean and standard deviation calculated in the training dataset. Unlike the standard batch normalization function, the $V_{th}$ is also taken into consideration. To account for the additional temporal domain, the mean and variance are calculated for each channel not only for all the spatial dimensions but also for the temporal dimension. To reduce computation complexity, we merge batch normalization with the pre-synaptic calculation described in~\eqref{eq:inp_current}. 

When we export parameters for hardware deployment, batch normalization is fused into weight and bias to eliminate overhead in inference as in 
\begin{equation}
\label{eq:w_fuse}
W_{fused} = \frac{V_{th}W}{\sqrt{Var[x]+\epsilon}}*\gamma 
\end{equation}
and
\begin{equation}
\label{eq:b_fuse}
B_{fused} = \frac{V_{th}(B-\mu)}{\sqrt{Var[x]+\epsilon}}*\gamma + \beta.  
\end{equation}

\subsection{Training}
\label{sec:training}

Spiking functions for neurons in SNNs are not directly differentiable. We implement a surrogate function to allow the gradient to propagate back through the neurons, as suggested by~\cite{shrestha_slayer_2018,wu_spatio-temporal_2018}. In this work, we use a square surrogate function defined by 
\begin{equation}
\label{eq:surrogate_fun}
\frac{\partial a}{\partial u} = \begin{cases}
1  &  \text{if} |u(t) - V_{th}| < 0.5 \\
0  &  \text{else}.
\end{cases}
\end{equation}
In the forward path, the spike function is governed by \eqref{eq:spike_fun}, while, in the backpropagation phase, the gradient is calculated by the surrogate function. We choose this surrogate function for its simplicity and good performance shown in other works~\cite{wu_spatio-temporal_2018}.

The training process is developed on top of the publicly available PyTorch implementation of the STBP backpropagation training method introduced in~\cite{wu_spatio-temporal_2018}. We modify it to support the proposed architecture incuding the reset-by-subtract scheme, trainable decay factor, quantization, hyperparameter sweeps, regularization as well as non-spiking output layers.


\begin{figure}[t]
  \centering\includegraphics{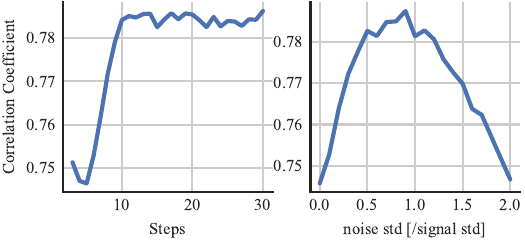}
 \caption{(Left) Time step sweep. The number of time steps for the network to unfold in the training phase. The time step of 10 is chosen as the accuracy does not increase with more time steps. (Right) Sweep strength of the noise. The noise is white Gaussian noise, whose standard deviation is set to a ratio relative to the mean standard deviation of signals from all channels. Best performance is achieved at 0.9.}
 \label{fig:noise_step_sweep}
\end{figure}

Fig.~\ref{fig:network_arch} shows the unfolded network during the training process. The STBP allows backpropagation through both temporal and spatial dimensions. The network is unfolded for ten time steps during the training process. This is equivalent to having a much deeper network, similar to the training process of RNNs. The time sequence length for training should be long enough to capture the relevant temporal dependencies; however, extending the number of time steps overly may lead to longer training duration and worsen the issue of vanishing or exploding gradients. We observed that the network does not need to unfold more than ten frames as it does not improve the accuracy, shown in Fig.~\ref{fig:noise_step_sweep}. 
Additionally, we discard the first two frames in the loss calculation because the network has not converged to a stable prediction, yet. The loss is then back-propagated through spatial and temporal dimensions in this unfolded network. 

We generate the training dataset by using a sliding window with a length of 10 time steps and an overlap of 9 time steps. These training samples are then shuffled during training.
Note that during the inference process, the network is not unfolded, and all neurons maintain and update their internal state across multiple inferences. One time step is used once per inference. This operating scenario mimics the situation where a real-time prediction task runs on data that are streamed continuously to the network.

We apply the AdamW optimizer~\cite{loshchilov_decoupled_2018} with a learning rate and a weight decay of $2\times10^{-3}$ and $1\times10^{-2}$, respectively. Learning rate decay is configured to decay by 1 decade every 20 epochs for better convergence. The batch size used during training is 128. The membrane threshold $V_{th}$ is set to $0.4$ which facilitates an optimal firing rate, thereby contributing to improved accuracy. These hyperparameters are determined by grid search. 

Similarly to~\cite{willsey_real-time_2021}, we use the time-integrated mean square error as the loss function during training, while the Pearson correlation coefficient and mean square error are used as the metric to evaluate the performance of the SNN, which are commonly adopted for neural decoding algorithm comparisons~\cite{nason_low-power_2020,willsey_real-time_2021}.

\subsection{Regularization}
\begin{figure}[t]
 \centering\includegraphics[width=0.9\linewidth]{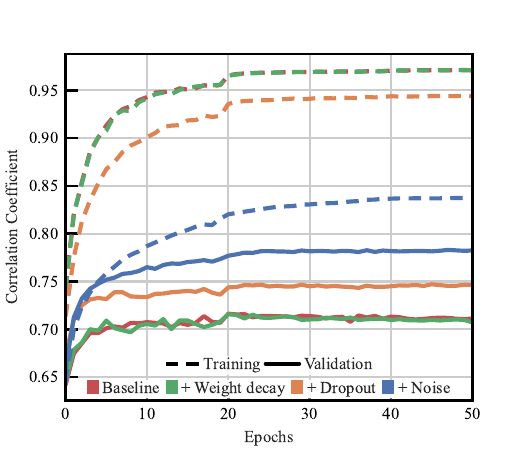}
 \caption{The correlation coefficients over epochs for the training set (dash line) and the validation set (solid line). The gap between the training line and the corresponding validation line shows the effectiveness of the regularisation methods. The baseline is the case when no regularisation method is applied. Weight decay does not help. Dropout and noise injection significantly alleviate the overfitting and improve accuracy.} 
 \label{fig:overfit}
\end{figure}

Overfitting is a common problem for neural networks that have numerous parameters and are trained with limited data. Our model also suffers from overfitting as observed in Fig.~\ref{fig:overfit}. 
We address this issue by three means. 
First, the weight decay function from the AdamW optimizer is applied. 
Second, dropout is implemented between each layer. It is performed for only spatial dimensions. At each time step, a new dropout mask is generated randomly. The dropout probability must be carefully calibrated. While low dropout probability provides insufficient regularization, excessively high probability can lead to the neglect of too many neurons during training, causing significant information loss. After a grid search, a dropout probability of 0.2 is chosen.

Third, white Gaussian noise is inserted before normalizing the input data to enhance the regularization effect. The inserted noise has a zero mean, and the standard deviation is set to 0.9 times the average standard deviation of the input features across all channels. While the noise helps to mitigate overfitting by introducing variability and uncertainty during training, too strong noise obscures the input feature and reduces accuracy. The standard deviation of 0.9 is determined as a result of a parameter sweep shown in Fig.~\ref{fig:noise_step_sweep}. The noise added to input data from all the channels is calculated based on the same criteria mentioned above. 

Noise injection not only adds randomness to the input data but also serves as a data augmentation method. As introduced in section~\ref{sec:training}, each training sample is prepared by using a sliding window of 10 time steps with 9 steps overlapped with the previous sample. Noise is added after these 10-step-long samples are made. This means that the same original data in different sliding windows will be slightly different due to the added noise, and this effectively augments the dataset to 10 times larger.

The impact after adding each regularisation scheme is shown in Fig.~\ref{fig:overfit}. Weight decay does not appear to have a noticeable effect. Dropout, on the other hand, reduces the disparity between the training and validation curves. The best accuracy is attained after applying noise injection, which shows the effectiveness of this approach in mitigating the overfitting issue.

\section{Deployment}
\label{sec:hardware}
\subsection{Quantization}
Quantization is an effective method for improving hardware efficiency. Decreasing the bit precision of the data and weights reduces the memory footprint and enables the usage of lighter integer computations instead of expensive floating-point computations. Quantization unleashes the full capability of the target MCU equipped with 8-bit SIMD instructions.

Before performing quantization, the batch normalization parameters are first fused into weights and biases. Thanks to the 1-bit information carrier in an SNN, no scaling is required between layers. We implement uniform quantization with a symmetrical range. This quantization is performed on a layer-by-layer basis, with distinct scaling factors computed for each individual layer. The scaling factor is primarily determined by the weights. The scaling factors of weight, bias, and threshold voltage are calculated by the bit-width and the maximum values of the weights in that layer as $scale = 2^{(bitwidth - 1) - 1}/W_{max}$.

The quantized weight, bias, and threshold voltage are generated by rounding the scaled values to the closest integer as $quant = round(value\cdot scale)$.  The membrane potential uses more bits than the weights do to provide numerical stability.

The decay factor ranges from 0 to 1 and is quantized differently. The scaling factor of $\lambda$ is only determined by the chosen bit-width. Unlike the weight and bias, which are applied once per inference for various inputs, the decay factor is repeatedly multiplied by the membrane potential in each forward pass, so the scaling factor for the quantized decay factor may impair numerical stability. 
To address this issue, we re-scale the multiplication result back through right-shift and truncation after performing multiplication with the scaled $\lambda$, formulated as $u^{l}(t) = (\lambda 2^{bw} (u^{l}(t-1) - s^{l-1}(t-1)V_{th})>>bw + i^{l}(t)$.

To compensate for the accuracy loss caused by quantization, quantization-aware training (QAT) is adopted. After regular training epochs, QAT is applied for 20 epochs. During QAT, a straight through estimator is used~\cite{gholami_survey_2021}. In the forward path, the simulated quantization is applied to mimic the effects of quantization, while in the back-propagation, the error is back-propagated with full precision, as the gradients for the piece-wise flat operator for quantization are almost zero everywhere~\cite{gholami_survey_2021}.

The neural decoder shows almost no loss even when the weights are quantized to 4 bits and decay factors are quantized to 3 bits. 
The model is exported after quantization and used for hardware deployment.

\subsection{Hardware Platform}
\begin{figure}[t]
    \centering
    \includegraphics{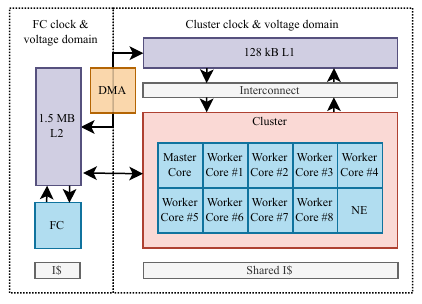}
    \caption{\textit{GreenWaves Technologies} GAP9 Microprocessor block diagram.}
    \label{fig:gap9_block_diagram}
\end{figure}

The network is deployed on a GAP9 hardware platform, a low-power 32-bit microcontroller (MCU), consisting of 10 RISC-V ISA cores as depicted in Fig.~\ref{fig:gap9_block_diagram}. A `Fabric Controller' core runs on startup and manages IO-related tasks, such as interfacing with peripherals and configuring operating voltages and frequencies. The platform is equipped with a compute cluster that allows task parallelization, consisting of a `master' core, eight worker cores, and an accelerator (NE). Each core on GAP9 supports an extended instruction set with support for zero-overhead hardware loops and various Single-Instruction-Multiple-Data (SIMD) operations for 8- and 16-bit data types.

The MCU uses a hierarchical memory architecture consisting entirely of SRAM, accessed through interleaved memory banks. The memory regions of relevance to this work are (1) 1.5 MB located near the Fabric Controller with a longer access time to the cluster known as \textit{L2}, (2) 128 kB located in the cluster with one cycle of access time known as \textit{L1}.

Considering the memory hierarchy, care must be taken to ensure that data is in a low-access cost memory region when it is needed for execution. To achieve this, a dedicated direct memory access (DMA) unit is equipped and can be used to move data between \textit{L2} and \textit{L1} at high bandwidth without interrupting execution. A single transfer command consists of source and destination addresses and the number of bytes to move.

\subsection{Data Movement \& Execution}

Eight cores in the compute cluster are used for most of the execution.
Between the execution of each layer, the master core is responsible for allocating and deallocating buffers, and configuring the tasks to be executed on each worker core. The model parameters and the membrane potentials are ordinarily stored in the \textit{L2} memory region. The weights for each layer are stored in a 2D array. The binary input spike vector determines which columns are used during the weight summation operation along each row of the weight matrix. This is illustrated in Fig.~\ref{fig:sparse-copy} where the weights are denoted by $W_{ij}$, and $i$ and $j$ represent the neurons in the current and previous layers, respectively. To decrease memory access times, the parameters and inputs are transferred to \textit{L1} for execution. 

However, it is not feasible to transfer the entire model from the \textit{L2} memory region to \textit{L1} due to the model size. 
Additionally, transferring the entire model is not optimal due to the latency and energy overhead. Therefore, we propose different data moving and execution strategies for different layers according to their characteristics.


\subsubsection{Layer 1}
\label{sec:snn_layer1}
The input to the first layer is an 8-bit normalized SBP feature vector. The data movement and execution scheme for layer 1 is similar to double buffering, where the data transfer runs concurrently with the model execution. Firstly, the input vector for the model is transferred to \textit{L1}, followed by the membrane potentials of all four layers. Then, the parameters for the first layer are split into two blocks of 128 rows, where execution can begin as soon as the first block has been transferred. Once the first block has finished executing and the parameters for the second block have arrived in \textit{L1}, the second block can be executed. This hides the latency associated with transferring the second half of layer 1's parameters, as demonstrated by the program profiler trace in Fig.~\ref{fig:trace}.

The blocks of parameters were configured to contain all parameters needed to execute a portion of the layer in a single contiguous region of memory. This allows all parameters for layer 1 to be moved into \textit{L1} with only two transfer commands, minimizing the cost of queuing a transfer. Each block contains 128 rows of the weight matrix $\mathbf{W}$ and the 128 corresponding bias and decay factors. The threshold voltages are read once from \textit{L2} at the start of each layer's execution by the cluster master core. The task is split row-wise as there are no data dependencies between rows. The computation is then distributed and parallelized on the 8 cluster cores, with each core processing 16 rows of each block in layer 1.

The worker cores perform the weight matrix multiplication step using a dot product SIMD operation available in each core on the GAP9 platform. This instruction computes the dot product between two 4-element 8-bit vectors, allowing four values from each row of the weight matrix to be processed per cycle per core. Once the result for a row has been calculated, the reset and decay factors are applied using 2-element 16-bit vector instructions.

\subsubsection{Layer 2, 3 \& 4}
\begin{figure}[t]
    \centering
    \includegraphics{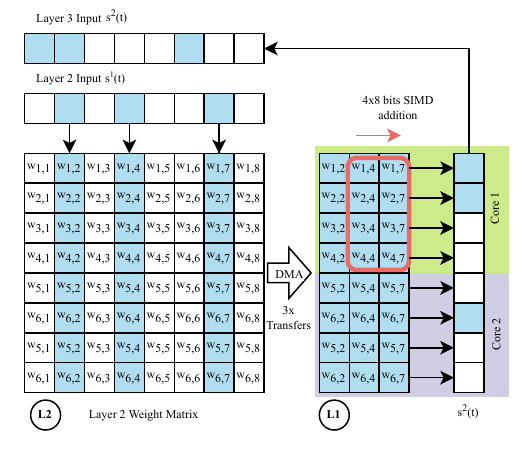}
    \caption{\textit{Sparse Copy} - The weight matrix calculation and initiation of transfers for the next layer are depicted in this diagram for a reduced weight matrix size. The columns required for computation are transferred from \textit{L2} to \textit{L1} using the DMA peripheral. The membrane potentials are then updated across multiple cores (two are shown here) using 4-element and 2-element SIMD. Once the spike conditions of the neurons in the layer are known, the process may repeat for the next layer.}
    \label{fig:sparse-copy}
\end{figure}

\begin{figure*}[t]
\centerline{\includegraphics[width=0.98\linewidth]{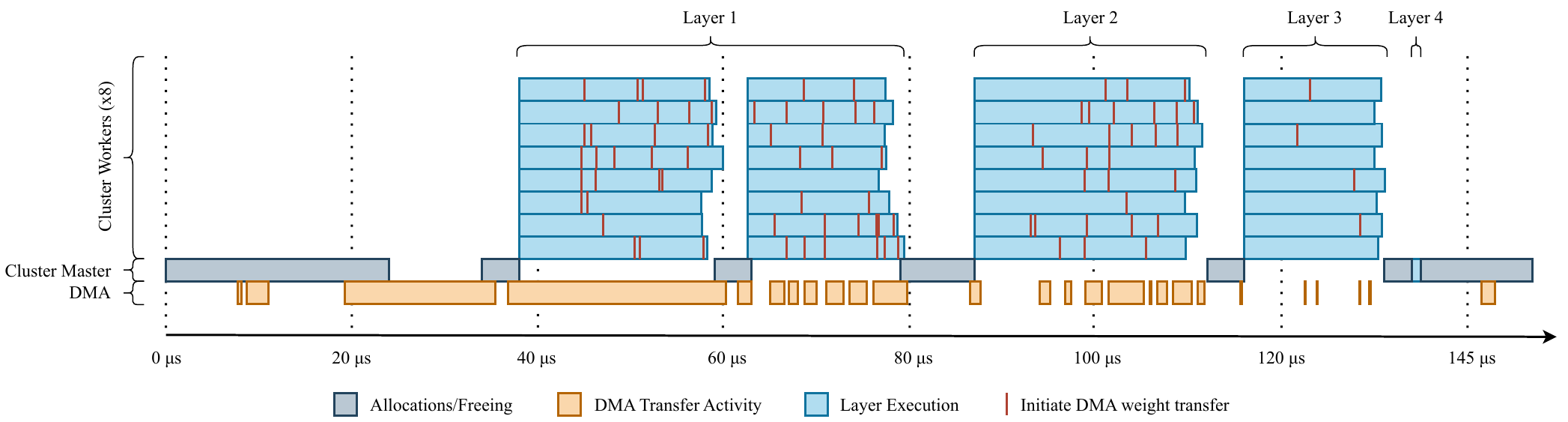}}
\caption{Profile trace for the sparse copy implementation running at 150 MHz depicting how DMA transfers are initiated for columns in subsequent layers. After initialization, the inputs and membrane potentials are transferred to \textit{L1}. Then the first half of layer 1's parameters are transferred before execution begins. Once the first block has finished executing and the DMA is finished transferring the second block, execution continues to the second half of layer 1. The DMA queue operations for sparse copy are shown in red. Many queue operations are omitted for the sake of clarity.}
\label{fig:trace}
\end{figure*}

In contrast to the first layer, each subsequent layer receives a binary vector as input. This is an important characteristic since it ensures that only the columns in the weight matrix corresponding to neurons fired in the preceding layer contribute to the update of the membrane potential. Thus, we propose the \textit{Sparse Copy} implementation, taking advantage of sparsity by only transferring the weights required for execution from \textit{L2} to \textit{L1} (as illustrated in Fig.~\ref{fig:sparse-copy}). This reduces energy consumption and execution time by decreasing the amount of data transferred. 
Instead of transferring data after a layer's execution is finished, the transfer is queued during layer execution to give the DMA more time to complete transfers. Once the membrane potential for a neuron is updated, the spike condition is checked, and the transfer of the corresponding weight column in the next layer is initiated. This allows the DMA peripheral to process many of the transfers while the cluster is still executing. This behavior is depicted in Fig.~\ref{fig:trace}. 
Two large buffers are used in \textit{L1}: one for reading during execution, the other for transferring new data. Due to the limited size of \textit{L1}, the size of these buffers is less than the total size of the layer 2 or layer 3 weight matrices. However, in practice, the number of neuron spikes is never large enough to result in a buffer overflow, thanks to the sparsity.
Once the transfer is queued, the destination address for the next transfer in the \textit{L1} weight buffer is incremented by 256 Bytes, the size of one column in the weight matrix. Besides saving the weights transfer, the sparse copy strategy also eases the write-back of the results. Saving the binary spikes is no longer required because the necessary DMA commands to obtain the relevant weights for the next layer are issued immediately after comparing the membrane potential to the threshold voltage.

The transfers were initiated using a DMA mode that allows cores to initiate transfers in chronological order, forcing subsequent transfers to wait. Because the transfer size is very small, i.e., 256 bytes, the time spent waiting is usually short. When multiple cores experience a spike in a short window, this latency accumulates and may affect execution time. This issue is further exacerbated for layers with a higher spike rate. 
An advantage of the proposed implementation is the capability of distributing the combined overhead of transfer initialization across all 8 cores rather than just the master core, as well as masking the transfer time behind the execution.

\begin{table}[b]
\begin{center}
\caption{Model Parameters and Associated Quantisation.}
\begin{tabular}{llcr} 
\hline
\textbf{Parameter} & \textbf{Layers} & \textbf{Bit Precision} & \textbf{Total size for all layers}\\
 \hline

$W$  & 1, 2, 3, 4 & 8 bits & 156,160 Bytes\\ 
$\lambda$ & 1, 2, 3, 4 & 16 bits & 1,540 Bytes\\
$B$ & 1, 2, 3, 4 & 16 bits & 1,540 Bytes\\
$V_{th}$  & 1, 2, 3 & 16 bits & 6 Bytes\\

$u$ & 1, 2, 3, 4 & 16 bits & 1,540 Bytes\\
\hline
\textbf{Total} & & & 160,786 Bytes\\
\hline
\end{tabular}
\label{tab:qparam}
\end{center}
\end{table}
As with layer 1, layers 2 and 3 are executed on all 8 cluster cores. Each core processes 32 of the 256 neurons in each layer. The program needs to keep track of the location to write the next column of the weight matrix, so there is a risk that two cores might initiate DMA transfers to the same memory address, in which case, one overwrites the other. This problem is solved using a hardware semaphore, which supports atomic integer increment and decrement. It is initialized to 0 before each layer executes. Whenever a core wants to transfer a column of weights, it loads and increments the semaphore value to determine the destination address in the \textit{L1} weight buffer. It can then safely queue a DMA transfer to this address from the column's memory address in \textit{L2}. It is not necessary for the weight matrix in \textit{L1} to maintain the original column order as in the weight matrix in \textit{L2}, the elements transferred to \textit{L1} only need to be summed in a row-wise manner.


As with Layer 1, SIMD operations are used to perform the weight matrix calculation. A 4-element 8-bit addition instruction allows four rows of each column in the weight matrix to be added per cycle, as shown in Fig.~\ref{fig:sparse-copy}. The quantization configuration used during training guarantees that the 8-bit signed addition will not over- or underflow. 

Finally, because the fourth layer consists of only two rows, it is executed on just the master core. Additionally, the layer 4 weights are padded to 16 bits to enable the use of 2-element 16-bit addition SIMD operations.

\section{Results and discussion}
\label{sec: result}
\subsection{Environment}
The model is constructed in Python with the Pytorch framework. The training and hyperparameter search are facilitated with Raytune framework to enable parallel training on multiple GPUs. The code for deployment to MCU was written in C and then compiled and deployed with GAP9 SDK Release v5.11.0. 

The algorithm is trained and evaluated on two datasets recorded from non-human primates while they were performing two-degree-of-freedom finger tasks. The datasets contain positions and velocities of two fingers, as well as the SBP feature from 96 channels. Dataset A, also used in~\cite{willsey_real-time_2021}, contains $\SI{817}{s}$ data. Dataset B is an open-source dataset, used also in~\cite{nason_real-time_2020}, containing $\SI{610}{s}$ data\footnote{Dataset is available for download: \url{https://deepblue.lib.umich.edu/data/concern/data_sets/0g354f51t}}. The SBP feature in the dataset is sampled at \SI{2}{kHz}. The SBP is first time-averaged in non-overlapping time bins before being fed into the proposed decoder. Time bin sizes are chosen to be $\SI{50}{ms}$ and $\SI{32}{ms}$ for dataset A and B, respectively, as introduced in~\cite{willsey_real-time_2021,nason_real-time_2020}. SBP features are standardized by removing means and scaling to a standard deviation of 1 before being fed into the SNN. Predicted velocity is also standardized with statistics from the training set. Standardized velocity is recovered to the original scale after inference. Hyperparameter optimization for the training parameters, unfolding time steps, and noise standard deviation is only conducted on dataset A.

The first $80\%$ of the data are used for training, and the remaining $20\%$ are for validation. The non-quantized model is trained for 60 epochs. The quantized SNN is trained with full precision for 30 epochs before QAT is enabled for another 20 epochs.  An inference is performed for every time frame to generate two-finger velocities in a streaming fashion.

\begin{figure}[t]
\centerline{\includegraphics[width=0.95\linewidth]{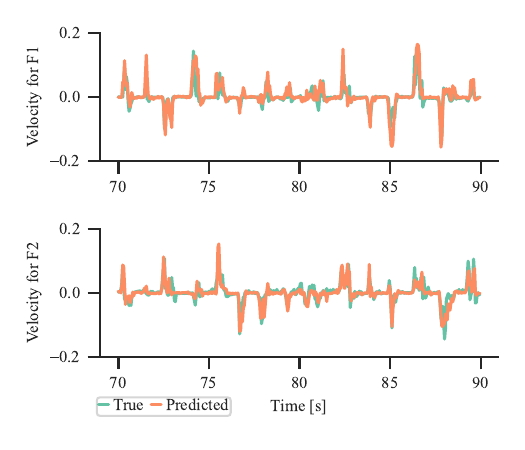}}
\caption{Predicted velocity versus true velocity for two fingers.}
\label{fig:velocity}
\end{figure}

\subsection{Memory footprint}
Although the weights can be quantized to 4 bits, they are padded to 8 bits for the hardware deployment to take advantage of the 8-bit SIMD operations of the MCU. The bit precision for different parameters and their memory footprint are presented in Table~\ref{tab:qparam}. Layer 4 is a non-spiking layer, so no threshold voltage is required. In total, the model requires 159,246 Bytes of storage for parameters or 160,786 Bytes, including the neuron membrane potentials.

\subsection{Accuracy}
The velocity predicted by the proposed SNN, and the true velocity of two fingers are plotted in Fig.~\ref{fig:velocity}. The accuracy is presented as the correlation coefficients between the predicted velocity and the recorded velocity in the validation set. The reported correlation coefficients are the mean values for two finger velocities. Table~\ref{tab:acc_comp} summarises the mean correlation and root mean square error (RMSE) over eight runs. 
To compare our results with previous work, we replicate the KF predictor~\cite{nason_real-time_2020} and the ANN predictor~\cite{willsey_real-time_2021} using the same parameters as the original papers. Our proposed work and the quantized model reach the highest correlation coefficients of 0.783 and 0.624 for datasets A and B, respectively.

\begin{figure*}[!t]
 \centering\includegraphics[width=\linewidth]{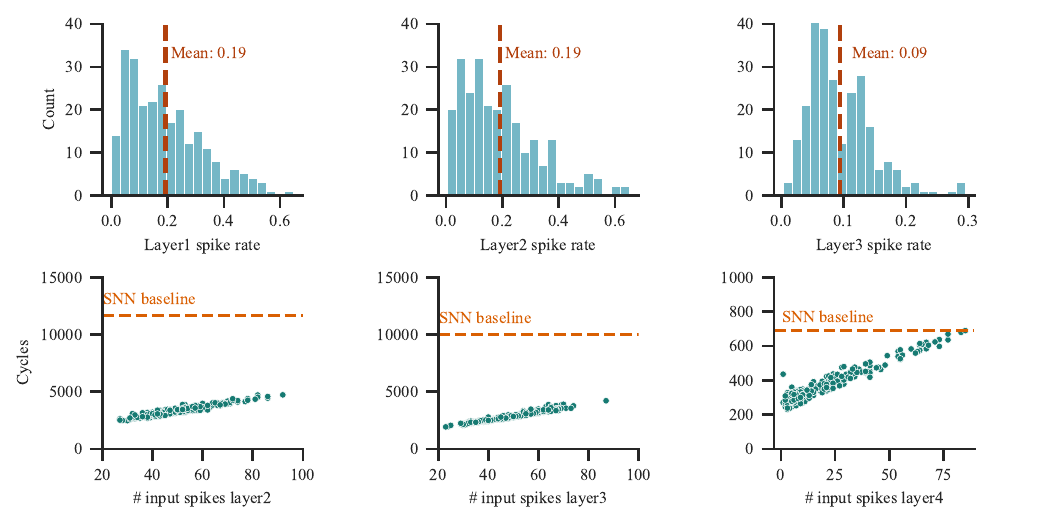}
 \caption{(Top) Spike rate distribution for three spiking layers. Each bar indicates the number of neurons exhibiting a certain spike rate. (Bottom) The number of cycles for the processing of a layer for different numbers of input spikes. Dots represent measured cycles for SNN sparse copy implementation. The number of cycles required for the SNN baseline implementation without using sparsity is illustrated by the orange lines}
 \label{fig:spike_rate_cycle}
\end{figure*}

\subsection{Computation complexity analysis}
The SNN's potential to achieve energy-efficient computation stems from the sparsity of the spikes. The spike rate for neurons and the average spike count in different layers are evaluated for the whole validation set of dataset A to quantify the sparsity in our network. The spike rate is calculated as the ratio of a neuron to fire over the total number of inferences. Fig.~\ref{fig:spike_rate_cycle} shows the histogram for the spike rate distribution on the three spiking layers. The average spike rates for the three layers are 19\%, 19\%, and 9\%, respectively. Out of the 770 neurons, only 120 of them fire in each inference step on average. Ideally, this high degree of sparsity significantly reduces the necessary number of memory accesses and computation operations. The required operations are summarised in Table~\ref{tab:comparison}.  
\begin{table}[t]
\begin{center}
\caption{Accuracy in terms of correlation coefficients and RMSE.}
\begin{threeparttable}
\begin{tabular}{lrrrr} 
\hline
\textbf{Decoder} & \textbf{Corr A} & \textbf{RMSE A} & \textbf{Corr B} & \textbf{RMSE B} \\
 \hline

KF ~\cite{nason_real-time_2020}  & 0.601$^\dagger$ & 0.026$^\dagger$& 0.459$^\dagger$ & 0.016$^\dagger$\\ 
ANN ~\cite{willsey_real-time_2022} &  0.732$^\dagger$ & 0.032$^\dagger$& 0.593$^\dagger$ & 0.017$^\dagger$\\
\textbf{Proposed SNN} & \textbf{0.783} & \textbf{0.022}& \textbf{0.624} & \textbf{0.014}\\
\textbf{Quantised SNN} & \textbf{0.782} & \textbf{0.021}& \textbf{0.627} & \textbf{0.014}\\

\hline
\end{tabular}
\label{tab:acc_comp}
\begin{tablenotes}
\footnotesize
\item $^\dagger$Reproduced results.
\end{tablenotes}
\end{threeparttable}
\end{center}
\end{table}

In conventional ANNs, the weighted sum computation for each input to a neuron requires a MAC operation. Whereas, in the proposed SNN, as the spike status is either 1 or 0, this process has been replaced by the add operations, which require much less power~\cite{horowitz_11_2014}. It is important to note that for a fair comparison between the proposed SNN and ANN, the SNN needs to update its membrane potentials, as described in \eqref{eq:mem_upd}, once per inference, which amounts to an additional MAC operation per each neuron. In this complexity analysis comparison, it is assumed that three memory loads and one store are required for each MAC operation, while for each addition, two loads and one store are required. 

As a reference point, we also include deployment of the same proposed SNN labeled as SNN baseline in our comparison, which does not exploit sparsity in any manner. Each layer in the SNN baseline implementation uses the double buffering approach described in section~\ref{sec:snn_layer1}.

Thanks to the high degree of sparsity, the proposed SNN requires up to one order of magnitude fewer operations and memory accesses than the ANN and the SNN baseline implementation.

\begin{table*}[!htbp]
\begin{center}
\caption{Performance and computation complexity analysis.}
\setlength{\tabcolsep}{10pt}
\begin{threeparttable}
\begin{tabular}{lrrrrrrr} 
\hline
& \multicolumn{4}{|c|}{Theoretical complexity analysis} & \multicolumn{3}{c|}{Measurement} \\
 &  \textbf{\# Parameters} & \textbf{MAC} & \textbf{ADD} & \textbf{Memory access}& \textbf{Avg. Power\tnote{a}}& \textbf{Energy / Inf.\tnote{b}} &\textbf{Time/ Inf.}\\
\textbf{Decoder} & \textbf{(K)} & \textbf{(K)}  & \textbf{(K)} & \textbf{(K)} & \textbf{(mW)} & \textbf{(uJ)} & \textbf{(ms)} \\
 \hline

ANN ~\cite{willsey_real-time_2022}  & 526 & 529 & - & 2116 & 0.60 & 10.69 & 0.66\\
SNN baseline  & 158 & 25 & 132 & 496 & 0.57 & 3.90 & 0.22\\
\textbf{SNN sparse copy} & \textbf{158} & \textbf{25} & \textbf{26} & \textbf{178} & \textbf{0.50} & \textbf{1.88} & \textbf{0.12}\\

\hline
\end{tabular}
\label{tab:comparison}
\begin{tablenotes}
\footnotesize
\item[a] Average power measured when the MCU is waked up every \SI{50}{\milli\second}, and after the inference, the MCU is put into sleep mode.
\item[b] Energy/inference measured when the MCU performs continuous inference without sleep to show the best achievable energy efficiency.
\end{tablenotes}
\end{threeparttable}
\end{center}
\end{table*}

\subsection{Cycle Count Comparison}
The preceding analysis examines the potential savings in SNN operations under ideal circumstances. However, it is important to note that there may be additional overhead in actual implementation. For example, addressing the irregular access pattern resulting from sparsity may require longer processing times. Therefore, we implement the decoder described in section~\ref{sec:hardware} targeting deployment on GAP9. 

\begin{figure}[t]
\centerline{\includegraphics[width=\linewidth]{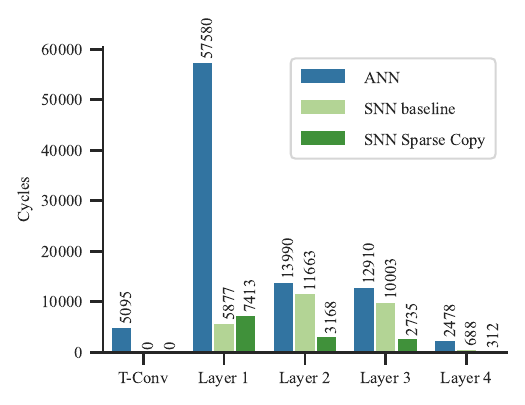}}   
\caption{Number of cycles used per layer for the ANN and SNN implementations on GAP9. The reported SNNs include an SNN baseline implementation without sparsity optimization, and an SNN effectively utilizes the sparsity noted as an SNN sparse copy. Note that the SNN does not have a temporal convolution layer, and the first fully connected layer of the ANN is 16 times larger than that of the SNN.}
\label{fig:cycles}
\end{figure}

First, we assess the number of cycles required for the inference to understand the latency of the decoder using the GAP9 SDK.  We compare the implementations in our work to a baseline ANN~\cite{willsey_real-time_2022} that is trained for the same task and deployed on the same hardware platform using the \textit{nntool}, which is an automatic neural network deployment tool provided by the SDK, with the accelerator enabled. 
The corresponding cycle counts measured for each layer are shown in Fig.~\ref{fig:cycles}. The cycle counts are averaged over 500 inferences from the validation dataset and are measured by finding the time taken from the end of execution of the previous layer to the end of execution of the current layer. The time for copying the SBP inputs and membrane potentials to \textit{L1} is not included in the individual layer cycle counts. The ANN takes many more cycles in the first layer due to the larger fully connected structure after the temporal convolution. All the implementations have the same sizes for the layer 2,3, and 4. For the SNN implementations, the input to layer 2, 3, and 4 are binary spikes. Therefore, the sparse copy implementation can leverage the sparsity and achieve the lowest cycle counts. The overhead of queuing DMA transfers in the sparse copy implementation for the weights of the subsequent layer is demonstrated in the increased execution time of Layer 1 of the sparse copy method over the SNN baseline implementation. Nevertheless, the gains made in subsequent layers justify this cost.

To further evaluate how well the sparsity is utilized, we also measure the number of cycles required for each layer when there are different numbers of spikes. The number of spikes is data-dependent. The distribution for spike counts observed while performing inference on the validation dataset is shown in Fig.~\ref{fig:spike_rate_cycle} for the sparse copy implementation. The cycle count for the sparse copy exhibits a strong linear relationship with the spike count, which is always significantly lower than that of the SNN baseline implementation in the measurement range. However, it is expected that at a higher spike rate, i.e., lower sparsity, the slope will change as the DMA peripheral becomes saturated and the bottleneck shifts away from execution towards data transfer.

\subsection{Power, energy, and latency measurement}
The measurements for power, energy, and latency are done on the physical GAP9 module.  The chip runs at \SI{150}{\mega\hertz}, \SI{0.65}{\volt}, for all measurements in order to maximize energy efficiency. The current consumption of the GAP9 chip is measured using a DC power analyzer, from which the power usage is then calculated.

The measurement results for the SNN baseline, sparse copy, and ANN implementations, taken over 3000 inferences, are shown in Table~\ref{tab:comparison}. The average energy consumption for the proposed implementation when running the inference continuously is \SI{1.88}{\micro\joule} per inference and the average latency is \SI{0.12}{\milli\second}, while the SNN baseline implementation takes \SI{3.90}{\micro\joule} per inference and \SI{0.22}{\milli\second} latency. The baseline ANN takes a much higher energy of \SI{10.69}{\micro\joule} per inference and \SI{0.66}{\milli\second}.

\begin{figure}[t]
\centerline{\includegraphics[width=\linewidth]{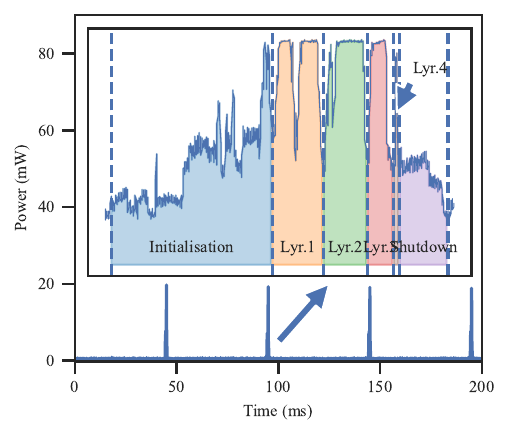}}
\caption{Power trace captured over four inferences of the SNN sparse copy implementation, running at \SI{150}{\mega\hertz}, \SI{0.65}{\volt}; including sleep between each inference. A zoomed-in capture of a single inference is shown in the top right.}
\label{fig:power}
\end{figure}

For dataset A, which is utilized in the power measurements, the sampling interval is \SI{50}{ms}. To save power and energy, the chip is configured to a `light sleep' mode between inferences. In this mode, the cluster and fabric controller are powered down, reducing consumption to $\approx$ \SI{395}{uW}.  However, \textit{L2} is the only retentive memory region during sleep, requiring the membrane potentials for each layer to be copied to \textit{L2} between each inference. A GPIO pad was configured as a `wake-up source' from sleep and triggered from an external waveform generator every \SI{50}{\milli\second}. The power trace for four inferences is shown in Fig.~\ref{fig:power}. When the time spent asleep is included, the average power usage is just \SI{0.50}{\milli\watt}. Due to the long sleep time in this setup, the power is dominated by the sleeping power. 

\subsection{Discussion}
The proposed implementation shows significantly improved performance, particularly in accuracy, energy efficiency, and latency for continuous measurement. This can be attributed to the following factors: 
1) The sparsity is efficiently utilized. The proposed implementation effectively reduces the number of memory accesses and executions. 2) The proposed decoder exhibits a high level of sparsity. Utilizing sparsity in MCU comes with the overhead of handling irregular patterns. This includes the need to determine which parameters to retrieve based on the output of each individual neuron. An excess of spikes could lead to congestion and degrade performance. However, the high sparsity in our network prevents us from these issues. 3) The MCU features a multi-core cluster with a DMA. The DMA allows us to hide the latency of fetching the irregularly positioned parameters by transferring data while executing. However, if more spikes occur, the DMA peripheral can be saturated, and the bottleneck may shift from execution to data transfer.
Therefore, it is essential to analyze the spike rate for each layer to develop effective strategies for data execution and transfer for the best efficiency.
4) The proposed SNN has a much higher number of connections compared to neurons. Despite the overhead of calculating each neuron's internal status, the overhead is negligible in our fully connected structure. In total, the SNN contains 156160 connections while having only 770 neurons. Therefore, the number of operations is dominated by the operations related to the connections between neurons instead of the neurons' internal states. 

Our work demonstrates the advantage of using SNNs compared to conventional methods in an application where low energy and power consumption is a critical requirement. Our optimized implementation and in-depth complexity analyses provide important insight for future work in automatizing SNN deployment on ultra-low-power hardware platforms.

\section{Conclusion}

We present a new SNN neural decoder to predict finger velocities along with its deployment method for implantable BMI and successfully demonstrate its capability in solving a real-world regression problem.
The proposed SNN is trained with STBP backpropagation enhanced by trainable decay factor, reset-by-subtract, and noise injection techniques to improve accuracy while keeping computation complexity low. The model is also fully quantized and deployed with optimization to efficiently utilize the sparsity on a general-purpose GAP9 MCU platform.

Compared to previous works, our SNN achieves the best correlation coefficient of 0.782 and 0.624 for the offline inference, while showing significantly less computation complexity, indicating potential in achieving energy-efficient hardware implementation. The deployed SNN on GAP9 achieves an average latency of \SI{0.12}{\milli\second} for each inference, which is 5.7X and 2.1X less than the baseline ANN and SNN baseline implementation without exploiting sparsity, respectively. 
The average power consumption, by duty cycling the MCU with sleep mode, is \SI{0.50}{\milli\watt}. The energy per inference assuming continuous inference without sleep is \SI{1.88}{\micro\joule}, which is 1.8X less than the SNN baseline implementation and 5.5X less than ANN. To the best of our knowledge, at the time of submission, this is the first sparsity-aware on-board demonstration showing that the SNN can be competitive in terms of latency and power consumption, even with general-purpose hardware platforms.

\section*{Acknowledgments}
We would like to thank Dr. Alfio Di Mauro for his advice on SNN training, Dr. Samuel R. Nason-Tomaszewski and Joseph Costello for their help on the datasets, and Cyrill Künzi for his contribution to the integration of the Raytune framework.












\bibliographystyle{IEEEtran}
\bibliography{main}

\end{document}